\documentclass[aps,pra,twocolumn,showpacs,superscriptaddress]{revtex4}
\usepackage{amsmath}
\usepackage{mathrsfs}
\usepackage{amssymb}
\usepackage{amsfonts}
\usepackage{graphicx}
\begin{document}

\title{Reduced fidelity susceptibility and its finite-size scaling
behaviors
in the Lipkin-Meshkov-Glick Model}

\author{Jian Ma}
\affiliation{Zhejiang Institute of Modern Physics, Department of
Physics, Zhejiang University, HangZhou 310027, P.R. China.}
\author{Lei Xu}
\affiliation{Zhejiang Institute of Modern Physics, Department of
Physics, Zhejiang University, HangZhou 310027, P.R. China.}
\author{Hengna Xiong}
\affiliation{Zhejiang Institute of Modern Physics, Department of
Physics, Zhejiang University, HangZhou 310027, P.R. China.}
\author{Xiaoguang Wang}\email{xgwang@zimp.zju.edu.cn}
\affiliation{Zhejiang Institute of Modern Physics, Department of
Physics, Zhejiang University, HangZhou 310027, P.R. China.}
\date{\today }
\begin{abstract}
We derive a general formula of the reduced fidelity susceptibility
when the reduced density matrix is $2\times2$ block-diagonal. By
using this result and the continuous unitary transformations, we
study finite-size scaling of the reduced fidelity susceptibility in
the Lipkin-Meshkov-Glick Model.  It is found that it can be used to
characterize quantum phase transitions, implying that we can extract
information of quantum phase transitions only from the fidelity of a
subsystem, which is of practical meaning in experiments.
\end{abstract}
 \maketitle

\section{Introduction}

During the past few years, some important concepts in quantum information
theory have been introduced to characterize quantum phase transitions
(QPTs). For example, entanglement, which is one of the central concepts in
quantum information theory, has been investigated extensively in QPTs in
various models, like Ising model \cite%
{Osterloh:Nature416,PhysRevA.69.022107,latorre:064101,barthel:220402}
and Lipkin-Meshkov-Glick (LMG) model \cite{PhysRevA.69.054101}.
Recently, fidelity, which is another important quantum information
concept, has also been applied in characterizing QPTs. The
introducing of fidelity\ in QPTs is natural
\cite{quan:140604,zanardi:031123,buonsante:110601,Zanardi:L02002,
cozzini:014439,you:022101,zhou:cond-mat/0701608,zhou:0704.2940,
zhou:0704.2945,venuti:095701,gu:0706.2495,Chen:arxiv/0706.0072,
Chen:arxiv/0708.3178,Kwok:arxiv/0710.2581v1,yang:180403,Zhao:arxiv/0803.0814, Chen:arxiv/0801.0020,Yang:arxiv/0803.1292}%
, since it's mathematically the overlap between two states, while
QPTs are just dramatic changes in ground-state properties. However,
the fidelity used in the study of QPTs depends computationally on an
arbitrary yet finite small change of the driving parameter. To
cancel the arbitrariness, Zanardi et al. introduced the Riemannian
metric tensor \cite{venuti:095701}, while You et al. suggested the
fidelity susceptibility \cite{you:022101}. The fidelity
susceptibility then becomes an effective tool to study critical
properties \cite{venuti:095701,gu:0706.2495} in many-body systems.

It's noticed that all the above works are concentrated on the
fidelity of the global ground states, and we may call this kind of
fidelity susceptibility the global fidelity susceptibility. However,
in experiments, one always probe the subsystem but not the whole
system for practical convenience. Here we use the reduced fidelity
\cite{gorin:quant-ph/0607050} (also called partial fidelity in
\cite{Paunkovic:arxiv/0708.3949,partialfide}) susceptibility (RFS),
which describes the fidelity susceptibility of a subsystem. In this
work, first we derive a general formula of the reduced fidelity
susceptibility when the reduced density matrix is $2\times2$
block-diagonal. Then, considering the LMG model, we show that the
RFS can be used to characterize QPTs, and find that the scaling
exponent is different from that of the global fidelity
susceptibility.

This paper is organized as follows. In Sec.~II, we briefly review
the concept of fidelity susceptibility, and give a general formula
of RFS for a special but interesting case that the density matrix is
$2\times 2$ block-diagonal. Then in Sec.~III, we introduce the LMG model \cite%
{Lipkin:188}. in the isotropic case, we find that the critical
behavior of RFS $\chi $ in response to magnetic transverse field $h$
as $(h_{c}-h) ^{-1}$ in thermodynamic limit. While in the
anisotropic case, by using the continuous unitary transformations
(CUTs) \cite{Wegner:Physik,Glazek:PhysRevD48,Glazek:PhysRevD49}, we
find that the maximum of $\chi $ over $h$ diverged as $N^{2/3}$ for
an $N$-spin system, and $\left| h_{c}-h\right| ^{-1} $ in
thermodynamic limit. Finally, we perform a numerical scaling
analysis, and the results are well consistent with our theoretical
ones.

\section{Reduced Fidelity Susceptibility}

We first give a brief review on the concept of fidelity
susceptibility. The
Hamiltonian of a quantum system undergoing QPTs can be written as%
\begin{equation}
H\left( h\right) =H_{0}+hH_{I},
\end{equation}%
where $H_{I}$ is supposed to be the driving term with control
parameter $h$. The global fidelity is defined as $F\left( h,\delta
\right) =\left\vert \langle \varphi _{0}\left( h\right) |\varphi
_{0}\left( h+\delta \right) \rangle \right\vert
$, where $|\varphi _{0}\left( h\right) \rangle $ is the ground state of $%
H\left( h\right) $, and $\delta $ is a small quantity. The reduced
fidelity is defined as the overlap between the reduced density matrix (RDM) $%
\rho \left( h\right) $ of the ground state $|\varphi _{0}\left( h\right)
\rangle $. In the follows, we take $\rho \equiv \rho \left( h\right) $ and $%
\tilde{\rho}\equiv \rho \left( h+\delta \right) $. Then the reduced
fidelity
is given by \cite{Uhlmann:rep-math-phys}%
\begin{equation}
F\left( h,\delta \right) =\text{tr}\sqrt{\rho ^{1/2}\tilde{\rho}\rho ^{1/2}}.
\end{equation}%
The corresponding fidelity susceptibility is defined as
\cite{zanardi:031123,you:022101}
\begin{equation}
\chi =\lim_{\delta \rightarrow 0}\frac{-2\ln F}{\delta ^{2}},
\label{def sus}
\end{equation}%
and then we could write $F\simeq 1-\chi \delta ^{2}/2$.

In this papar we consider that the RDM is block-diagonal,
\begin{equation}
\rho =\bigoplus\limits_{i=1}^{n}\varrho _{i},  \label{general rho}
\end{equation}%
where $\varrho _{i}$'s are $2\times 2$ semi-positive definite
Hermitian matrices, since $\rho $ is a density matrix. Now we
introduce some useful formulas at first. Let $A$ and $B$ are
arbitrary $2\times 2$ semi-positive definite
matrices, then we have%
\begin{equation}
\text{tr}\sqrt{A^{1/2}BA^{1/2}}=\sqrt{\text{tr}\left( AB\right) +2\sqrt{\det
\left( AB\right) }},  \label{tr_2body}
\end{equation}%
and if $A=B$, it becomes
\begin{equation}
\text{tr}\left( A^{2}\right) =\left( \text{tr}A\right) ^{2}-2\det A.
\label{tr_formula1}
\end{equation}%
Take derivations of the above equation with respect to some variable
$h$, we get
\begin{align}
\text{tr}\left( AA^{\prime }\right)  &=\text{tr}A\text{tr}A^{\prime
}-\partial _{h}\left( \det A\right) ,  \label{tr_formula2} \\
\text{tr}\left( AA^{\prime \prime }\right)
&=\text{tr}A\text{tr}A^{\prime \prime }-\partial _{h}^{2}\left( \det
A\right) +2\det A^{\prime }, \label{tr_formula3}
\end{align}%
where $A^{\prime }\equiv \partial _{h}A$, $A^{\prime\prime }\equiv
\partial _{h}^2A$ and $\partial _{h}$tr$\left(
A\right) =$tr$\left( A^{\prime }\right) $. Now the fidelity can be written as%
\begin{align}
F& =\sum_{i=1}^{n}\text{tr}\sqrt{\varrho _{i}^{1/2}\tilde{\varrho}%
_{i}\varrho _{i}^{1/2}}  \notag \\
& =\sum_{i=1}^{n}\sqrt{\text{tr}\varrho _{i}\tilde{\varrho}_{i}+2\sqrt{\det
\varrho _{i}\tilde{\varrho}_{i}}},  \label{fidelity}
\end{align}%
and recall that $F\simeq 1-\chi \delta ^{2}/2$, the susceptibility
$\chi =\sum_{i=1}^{n}\chi _{i}$, with $\chi _{i}$ corresponds to the
`susceptibility' of the $i$-th block in Eq.~(\ref{general rho}). To
obtain the susceptibility, we should expand the fidelity with
respect to $\delta $, and for $\tilde{\varrho}_{i}$ $\simeq $
$\varrho _{i}\left( h\right) +\varrho _{i}^{\prime }\left( h\right)
\delta +\delta ^{2}\varrho
_{i}^{\prime \prime }\left( h\right) /2+O\left( h^3\right) $, we have%
\begin{gather}
\left\{ \begin{aligned}
&\text{tr}\left(\varrho\tilde{\varrho}\right)\simeq\text{tr}\left(
\varrho^{2}\right) +\text{tr}\left( \varrho\varrho^{\prime}\right)
\delta+\frac{\delta^{2}}{2}\text{tr}\left(
\varrho\varrho^{\prime\prime }\right) ,\\
&\det\tilde{\varrho}\simeq \det\varrho+\partial_{h}\left(
\det\varrho\right) \delta+\frac{\delta^{2}}{2}\partial_{h}^{2}\left(
\det\varrho\right), \end{aligned}\right.   \label{eq:3}
\end{gather}%
here we omit the subscript $i$ for convenience.

In the case that $\det \varrho \neq 0$, we have tr$\varrho \neq 0$
since $\varrho $ is
semi-positive definite. Then we get%
\begin{align}
\sqrt{\det \left( \varrho \tilde{\varrho}\right) }& \simeq \det \varrho +%
\frac{\delta }{2}\partial _{h}\det \varrho   \notag \\
& +\frac{\delta ^{2}}{4}\left[ \partial _{h}^{2}\det \varrho -\frac{\left(
\partial _{h}\det \varrho \right) ^{2}}{2\det \varrho }\right] .
\end{align}%
Take the above expression into Eq.~(\ref{fidelity}) and with the help of
Eqs.~(\ref{tr_formula1}), (\ref{tr_formula2}) and (\ref{tr_formula3}) we
obtain
\begin{align}
& \text{tr}\sqrt{\varrho ^{1/2}\tilde{\varrho}\varrho ^{1/2}}\simeq \text{tr}%
\varrho +\frac{\delta }{2}\text{tr}\varrho ^{\prime }+\frac{\delta ^{2}}{4}%
\text{tr}\varrho ^{\prime \prime }  \notag \\
& +\frac{\delta ^{2}}{8\text{tr}\varrho }\left\{ 4\det \varrho ^{\prime
}-\left( \text{tr}\varrho ^{\prime }\right) ^{2}-\frac{\left[ \partial
_{h}\det \left( \varrho \right) \right] ^{2}}{\det \left( \varrho \right) }%
\right\} .  \label{fide1}
\end{align}%
If $\det \varrho =0$ but tr$\varrho \neq 0$, \ we have $\det \left( \varrho
\tilde{\varrho}\right) =0$. Moreover, since $\varrho $ is positive
semi-definite, zero is the lower bound of $\det \varrho $, which requires $%
\partial _{h}\det \varrho =0$ and $\partial _{h}^{2}\det \varrho >0$.
Thus we have
\begin{align}
\text{tr}\left( \varrho \tilde{\varrho}\right) & =\left( \text{tr}\varrho
\right) ^{2}+\text{tr}\varrho \text{tr}\varrho ^{\prime }\delta   \notag \\
& +\frac{\delta ^{2}}{2}\left[ \text{tr}\varrho \text{tr}\varrho ^{\prime
\prime }-\partial _{h}^{2}\left( \det \varrho \right) +2\det \varrho
^{\prime }\right] ,
\end{align}%
and%
\begin{align}
\text{tr}\sqrt{\varrho ^{1/2}\tilde{\varrho}\varrho ^{1/2}}& \simeq \text{tr}%
\varrho +\frac{\delta }{2}\text{tr}\varrho ^{\prime }+\frac{\delta ^{2}}{4}%
\text{tr}\varrho ^{\prime \prime }  \notag \\
& +\frac{\delta ^{2}}{8\text{tr}\varrho }\left[ 4\det \varrho
^{\prime }-\left( \text{tr}\varrho ^{\prime }\right) ^{2}-2\partial
_{h}^{2}\left( \det \varrho \right) \right] . \label{fide2}
\end{align}%
In the last case that tr$\varrho =0$, $\varrho $ is equivalent to a
zero
matrix, since $\varrho $ is Hermitian. Then tr$\left( \varrho \tilde{\varrho}%
\right) =\sqrt{\det \left( \varrho \tilde{\varrho}\right) }=0$, and $F=0$.

Conclude the above three cases, we get the `susceptibility' for
block $\varrho_{i}$ as
\begin{widetext}
\begin{equation} \label{gen_suses}
\chi_i=\left\{
\begin{split}
&\frac{1}{4\text{tr}\varrho_{i}}\left\{  \left(  \text{tr}%
\varrho_{i}^{\prime}\right)
^{2}-4\det\varrho_{i}^{\prime}+\frac{\left[
\partial_{h}\det\left(  \varrho_{i}\right)  \right]  ^{2}}{\det\left(
\varrho_{i}\right)  }\right\}\quad&\text{for}\quad&\text{tr}\varrho_{i}\neq0,\det\varrho_{i}\neq0,\\
&\frac{1}{4\text{tr}\varrho_{i}}\left[  \left(  \text{tr}\varrho_{i}%
^{\prime}\right)
^{2}-4\det\varrho_{i}^{\prime}+2\partial_{h}^{2}\left(
\det\varrho_{i}\right)  \right]\quad&\text{for}\quad&\text{tr}\varrho_{i}\neq0,\det\varrho_{i}=0,\\
&0\quad&\text{for}\quad&\text{tr}\varrho_{i}=0,
\end{split}
\right.
\end{equation}
\end{widetext}
where the terms of $\delta\text{tr}\varrho'/2$ and
$\delta^2\text{tr}\varrho''/4$ in Eqs.~(\ref{fide1}) and
(\ref{fide2}) are canceled in the final expression of the fidelity,
due to tr$\left( \rho \right) \equiv1$, and tr$\left( \rho ^{\prime
}\right) =\text{tr}\left( \rho ^{\prime \prime }\right) =0$.

Finally, we consider a more special case that $\rho $ is diagonal,
the
then susceptibility is obtained readily%
\begin{equation}
\chi =\sum_{i=1}^{n}\frac{\left( \lambda _{i}^{\prime }\right) ^{2}}{%
4\lambda _{i}},  \label{diag}
\end{equation}%
where $\lambda _{i}$'s are the nonzero diagonal terms.

\section{The LMG Model and its scaling exponents of RFS}

\subsection{The LMG model and RFS}

The LMG model was introduced in nuclear physics to describe mutually
interacting spin-1/2 particles, embedded in a transverse magnetic field. In
the thermodynamic limit, it undergoes a QPT that is described by the mean
field analysis \cite{PhysRevLett.49.478}. Recently the finite-size scaling
was studied by the $1/N$ expansion in the Holstein-Primakoff single boson
representation \cite{PhysRevB.28.3955} and by the CUTs \cite%
{PhysRevLett.93.237204,dusuel:224420}. The Hamiltonian of the LMG model reads%
\begin{align}
H=& H_{0}+hH_{I}  \notag \\
=& -\frac{\lambda }{N}\left( 1+\gamma \right) \left( \mathbf{S}%
^{2}-S_{z}^{2}-N/2\right)  \notag \\
& -\frac{\lambda }{2N}\left( 1-\gamma \right) \left(
S_{+}^{2}+S_{-}^{2}\right) -2hS_{z},
\end{align}%
where $S_{\alpha }=\sum_{i}\sigma _{i\alpha }/2$, with $\sigma _{\alpha
}\left( \alpha =x,y,z\right) $ the Pauli matrices, and $S_{\pm }=S_{x}\pm
iS_{y}$. The prefactor $1/N$ ensures finite energy per spin in the
thermodynamic limit. In the context, we set the parameters: $\lambda =1$, $%
\left\vert \gamma \right\vert \leq 1$, $h\geq 0$. We take $h\geq 0$ as the
spectrum is invariant under the transformation $h\leftrightarrow -h$. In
addition, we only consider the maximum spin sector $S=N/2$ in which the
lowest energy state lies.

Now we consider a 2-body RDM of the LMG model \cite{XWang:EPJD/18}
\begin{equation}
\rho _{ij}=%
\begin{pmatrix}
v_{+} & 0 & 0 & u \\
0 & y & y & 0 \\
0 & y & y & 0 \\
u & 0 & 0 & v_{-}%
\end{pmatrix}%
,  \label{reduced density matrix}
\end{equation}%
in the standard basis $\left\{ |00\rangle ,|01\rangle ,|10\rangle
,|11\rangle \right\} $, where $\sigma_{z}|0\rangle=-|0\rangle$ and $%
\sigma_{z}|1\rangle=|1\rangle$, while the nonzero matrix elements reads%
\begin{align}
v_{\pm }& =\frac{N^{2}-2N+4\left\langle S_{z}^{2}\right\rangle \pm
4\left\langle S_{z}\right\rangle \left( N-1\right) }{4N\left( N-1\right) },
\notag \\
y& =\frac{N^{2}-4\left\langle S_{z}^{2}\right\rangle }{4N\left( N-1\right) }%
, \quad u =\frac{\left\langle S_{x}^{2}-S_{y}^{2}\right\rangle
}{N\left( N-1\right) },  \label{matrix elements}
\end{align}%
where $\left[ A,B\right] _{+}=AB+BA$ is the anti-commutator for
operators $A$ and $B$. The zero elements of $\rho _{ij}$ result from
the fact that the
total spin and the parity are conserved quantities, i.e.,%
\begin{equation}
\left[ H,S^{2}\right] =\left[ H,\prod_{i=1}^{N}\sigma _{iz}\right] =0.
\end{equation}%
It's noticed that $\rho _{ij}$ is actually block-diagonal in the rearranged
basis $\{|00\rangle ,|11\rangle ,|01\rangle ,|10\rangle\}$, and the two
blocks are
\begin{equation}
\varrho _{1}=%
\begin{pmatrix}
v_{+} & u \\
u & v_{-}%
\end{pmatrix}%
,\varrho _{2}=%
\begin{pmatrix}
y & y \\
y & y%
\end{pmatrix}%
.
\end{equation}%
With the help of Eq.~(\ref{gen_suses}), we can give the RFS explicitly%
\begin{eqnarray}
\chi &=&\frac{y^{\prime 2}}{2y}+\frac{1}{4\left( v_{+}+v_{-}\right) }{\biggl[%
}\left( v_{+}^{\prime }-v_{-}^{\prime }\right) ^{2}+4u^{\prime 2}  \notag \\
&&+\frac{\left( v_{+}^{\prime }v_{-}+v_{+}v_{-}^{\prime }-2u^{\prime
}u\right) ^{2}}{\left( v_{+}v_{-}-u^{2}\right) }{\biggl],}  \label{gen_sus}
\end{eqnarray}%
here we consider the case that $\det\varrho_1\neq0$, and the
following computations are based on the above formula.

\subsection{The isotropic case}

Firstly, we consider the isotropic case, $\gamma =1$, and the Hamiltonian
reads
\begin{equation}
H=-\frac{2}{N}\left( \mathbf{S}^{2}-S_{z}^{2}-N/2\right) -2hS_{z},
\end{equation}%
which is diagonal in the standard eigenbasis $\left\{ |S,M\rangle \right\} $
of $\mathbf{S}^2$ and $S_{z}$. For $S=N/2$ the eigenstates are
\begin{equation}
E\left( M,h\right) =\frac{2}{N}\left( M-\frac{hN}{2}\right) ^{2}-\frac{N}{2}%
\left( 1+h^{2}\right) ,
\end{equation}%
and the ground state is readily obtained when
\begin{equation}  \label{eq:2}
M_0=\left\{ \begin{aligned}
&N/2 &\text{for}\quad&h\ge 1,\\
&N/2-R\left[N(1-h)/2\right] &\text{for}\quad&0\le h<1,
\end{aligned} \right.
\end{equation}
where $R(x)\equiv\text{round}(x)$. Then one can see level crossings
exist at $h=h_{j}$, where $h_{j}=1-\left( 2j+1\right) /N$, between
the two states $|N/2,N/2-j\rangle $ and $|N/2,N/2-j-1\rangle $. In
the thermodynamic limit, these critical points form a region of
criticality.

The elements of the RDM in ground state are readily obtained as
\begin{align}
v_{\pm}& =\frac{\left( N\pm2M_{0}\right) \left( N-2\pm2M_{0}\right) }{%
4N\left( N-1\right) },  \notag \\
y& =\frac{\left( N^{2}-4M_{0}\right) }{4N\left( N-1\right) },\quad
u=0.
\end{align}%
As $N$ is very large, $M_{0}\left( h<1\right) \simeq hN/2$. With Eq.~(\ref{gen_sus}), we obtain the susceptibility in thermodynamic limit%
\begin{equation}
\lim_{N\rightarrow \infty }\chi \left( h>1-\frac{1}{N}\right) \simeq \frac{1%
}{2\left( 1-h^{2}\right) }.
\end{equation}%
Obviously, the asymptotic behavior of $\chi $ as $h\rightarrow 1$ is $%
1/\left( 1-h\right) $. However, there is no QPT in its symmetric phase $%
h>1 $, because the ground state is independent of $h$.

\subsection{The anisotropic case}

\subsubsection{Spin expectation values}

\begin{figure}[tbp]
\begin{center}
\includegraphics[
height=2.5688in, width=2.8in ]{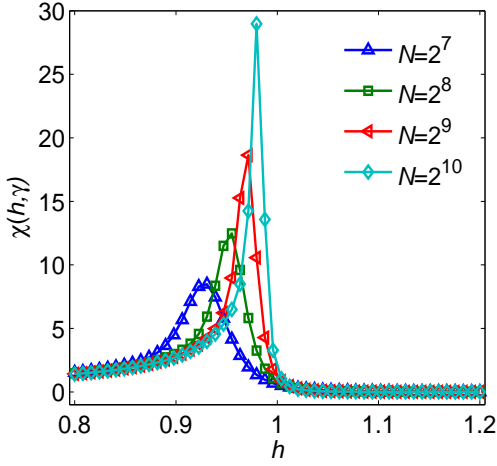}
\end{center}
\caption{Fidelity susceptibility $\protect\chi $ as a function of $h$ with
various system size $N=2^{7},2^{8},2^{9},2^{10}$. The positions of their
peaks approach to the critical point $h_{c}=1$ .}
\label{sus_h}
\end{figure}

Next we consider the anisotropic case, and the numerical results of the RFS
as a function of $h$ are shown in Fig.~(\ref{sus_h}). We adopt the $1/N$
expansion method with CUTs that was used extensively by Dusuel and Vidal
\cite{PhysRevLett.93.237204,dusuel:224420}, which corresponds to the large $%
N $ limit. While the Holestein-Primakoff method is not suitable for our task
since it could only give a first order correction in a $1/N$ expansion.

Here we firstly recall the CUTs introduced by Wegner \cite{Wegner:Physik}
and independently by Glazek and Wilson \cite%
{Glazek:PhysRevD48,Glazek:PhysRevD49}. For a pedagogical
introduction to this technique, one can see
\cite{dusuel:J.Phys.A37}. The main idea of CUTs is to diagonalize
the Hamiltonian in a continuous way starting from the original
Hamiltonian $H=H\left( l=0\right) $. A flowing Hamiltonian is then defined by%
\begin{equation}
H\left( l\right) =U^{\dagger }\left( l\right) H\left( 0\right) U\left(
l\right) ,  \label{flow eqn}
\end{equation}%
where $U\left( l\right) $ is unitary and $l$ is a scaling parameter such
that $H\left( l=\infty \right) $ is diagonal. A derivation of the Eq.~(\ref%
{flow eqn}) with respect to $l$ yields the flow equation%
\begin{equation}
\partial _{l}H\left( l\right) =\left[ \eta \left( l\right) ,H\left( l\right) %
\right] ,\label{flow_eqn2}
\end{equation}%
where $\eta \left( l\right) =-U^{\dagger }\partial _{l}U$ is an
anti-Hermitian generator. To obtain the expectation value of any
operator $\Omega $ on an eigenstate $|\psi \rangle $ of $H$, one
should follow the flow of the operator $\Omega \left( l\right) =$
$U^{\dagger }\left( l\right) H\left( 0\right) U\left( l\right) $, by
solving Eq.~(\ref{flow_eqn2}). Fortunately the results of the spin
expectation values have been obtained by Dusuel and Vidal in
\cite{PhysRevLett.93.237204,dusuel:224420}, and here we'll compute
the scaling behavior of the derivatives of these values.

Firstly, we consider the system size $N$ is very large, and the matrix
elements are rewritten as
\begin{eqnarray}
v_{\pm } &=&\frac{1}{4}+\frac{\left\langle S_{z}^{2}\right\rangle }{N^{2}}%
\pm \frac{\langle S_{z}\rangle }{N},  \notag \\
y &=&\frac{1}{4}-\frac{\left\langle S_{z}^{2}\right\rangle }{N^{2}},\quad u =%
\frac{\left\langle S_{x}^{2}\rangle -\langle S_{y}^{2}\right\rangle }{N^{2}}.
\label{mat_elem}
\end{eqnarray}%
The spin expectation values appeared in the above expressions can be solved
by the CUTs with $1/N$ expansion. For symmetry phase ($h>1$), we have%
\begin{widetext}
\begin{align}
\frac{2\langle S_{z}\rangle }{N}=& 1+\frac{1}{N}\left( \frac{P_{z}^{\left(
1\right) }}{G^{1/2}}+1\right) +\frac{\left( 1-\gamma \right) ^{2}}{N^{2}}%
\left( \frac{P_{z}^{\left( 2\right) }}{G^{2}}+\frac{Q_{z}^{\left( 2\right) }%
}{G^{3/2}}\right) +\frac{\left( 1-\gamma \right) ^{2}}{N^{3}}\left( \frac{%
P_{z}^{\left( 3\right) }}{G^{7/2}}+\frac{Q_{z}^{\left( 3\right) }}{G^{3}}%
\right) +O\left( \frac{1}{N^{4}}\right) , \nonumber\\
\frac{4\langle S_{x}^{2}\rangle }{N^{2}}=& \left( h-\gamma \right)
\left\{ \frac{1}{NG^{1/2}}+\frac{1}{N^{2}}\left(
\frac{P_{xx}^{\left(
2\right) }}{G^{2}}+\frac{Q_{xx}^{\left( 2\right) }}{G^{3/2}}\right) +\frac{1%
}{N^{3}}\left( \frac{P_{xx}^{\left( 3\right) }}{G^{7/2}}+\frac{%
Q_{xx}^{\left( 3\right) }}{G^{3}}\right) \right\} +O\left( \frac{1}{N^{4}}%
\right) , \nonumber \\
\frac{4\langle S_{y}^{2}\rangle }{N^{2}}=& \frac{1}{h-\gamma }\left\{ \frac{%
G^{1/2}}{N}+\frac{1}{N^{2}}\left( \frac{P_{yy}^{\left( 2\right) }}{G}+\frac{%
Q_{yy}^{\left( 2\right) }}{G^{1/2}}\right) +\frac{1}{N^{3}}\left( \frac{%
P_{yy}^{\left( 3\right) }}{G^{5/2}}+\frac{Q_{yy}^{\left( 3\right) }}{G^{2}}%
\right) \right\} +O\left( \frac{1}{N^{4}}\right) ,  \nonumber \\
\frac{4\langle S_{z}^{2}\rangle }{N^{2}}=& 1+\frac{1}{N}\left( \frac{%
P_{zz}^{\left( 1\right) }}{G^{1/2}}+2\right) +\frac{1}{N^{2}}\left( \frac{%
P_{zz}^{\left( 2\right) }}{G^{2}}+\frac{Q_{zz}^{\left( 2\right) }}{G^{3/2}}%
\right) +\frac{\left( 1-\gamma \right) ^{2}}{N^{3}}\left( \frac{%
P_{zz}^{\left( 3\right) }}{G^{7/2}}+\frac{Q_{zz}^{\left( 3\right) }}{G^{3}}%
\right) +O\left( \frac{1}{N^{4}}\right) ,  \label{spinval}
\end{align}%
\end{widetext}
where $G\equiv G\left( h,\gamma \right) =\left( h-1\right) \left( h-\gamma
\right) $. Here we do not present $P_{\xi }^{\left( i\right) }\equiv P_{\xi
}^{\left( i\right) }\left( h,\gamma \right) $ and $Q_{\xi }^{\left( i\right)
}\equiv Q_{\xi }^{\left( i\right) }\left( h,\gamma \right) $ ($i=1,2,3$ and $%
\xi =z,xx,yy,zz$), which are polynomials of $h$ and $\gamma $,
whereas of little meaning for computing the scaling exponents. For
more details, you can refer to the appendix part of
\cite{dusuel:224420}. It's noticed that, the above expressions can
be written in the form
\begin{equation}
\Phi _{N}\left( h,\gamma \right) =\Phi _{N}^{\text{reg}}\left( h,\gamma
\right) +\Phi _{N}^{\text{sing}}\left( h,\gamma \right) ,
\label{regular and singular form}
\end{equation}%
where the superscripts `reg' and `sing' stand for regular and
singular
respectively. A nonsingular contribution is understood to be a function of $%
h $ which is nonsingular at $h=1$, as well as all its derivatives. Take $%
2\langle S_{z}\rangle /N$ for example, the regular part is $1+1/N$ and the
remaining forms the singular part. As $h$ approaches to $1$, the terms
involving $Q_{\xi }^{\left( i\right) }$'s are small compared to the terms
involving $P_{\xi }^{\left( i\right) }$'s by a factor $G\left( h,\gamma
\right) $, hence we could only consider the terms involving $P_{\xi
}^{\left( i\right) }$'s.

\subsubsection{Finite-size scaling}

Here we show how to derive the finite-size scaling exponents of the spin
expectation values and their derivatives, and take $2\langle S_{z}\rangle/N$
for example,
\begin{eqnarray}
\frac{2\langle S_{z}\rangle }{N} &=&1+\frac{1}{N}+\frac{1}{NG^{1/2}}{\biggl
\{}P_{z}^{\left( 1\right) }+\frac{\left( 1-\gamma \right) ^{2}P_{z}^{\left(
2\right) }}{NG^{3/2}}  \notag \\
&&+\frac{\left( 1-\gamma \right) ^{2}P_{z}^{\left( 3\right) }}{\left(
NG^{3/2}\right) ^{2}}+O\bigg{(} \frac{1}{\left[ NG^{3/2}\right] ^{3}}\bigg{)}
{\biggl \},}
\end{eqnarray}%
where the singular part (terms after $1+1/N$) can be written in the
form
\begin{equation}
\left( \frac{2\langle S_{z}\rangle }{N}\right) ^{\text{sing}}\simeq \frac{1}{%
NG\left( h,\gamma \right) ^{1/2}}\mathscr{F}_{S_z}\left[ NG\left( h,\gamma
\right) ^{3/2},\gamma \right] ,  \label{scaling form of the
singular part}
\end{equation}%
where $\mathscr{F}_{\Phi}$ ($\Phi=S_{z},S_{x}^2,S_{y}^2,S_{z}^2$) is a
scaling function for these spin expectation values. While in fact that there
can be no singularity in any physical quantity in a finite-size system, and
the critical point $h_{c}=1$ only for thermodynamic limit $N\rightarrow
\infty $. This implies that the singularity of $G\left( h,\gamma \right)
^{-1/2}$ has to be canceled by the one of $\mathscr{F}_{S_z}\left[ NG\left(
h,\gamma \right) ^{3/2},\gamma \right] $. Thus one must have $\mathscr{F}%
_{S_z}\left( x,\gamma \right) \sim x^{-1/3}$, which in turn implies the
following finite size scaling:
\begin{equation}
\frac{2\langle S_{z}\rangle }{N}\bigg{|}_{h=1}\sim \frac{a_{z}^{\left(
0\right) }}{N^{2/3}},  \label{singular form}
\end{equation}

\begin{figure}[tbp]
\begin{center}
\includegraphics[
height=2.5613in, width=2.8in ]{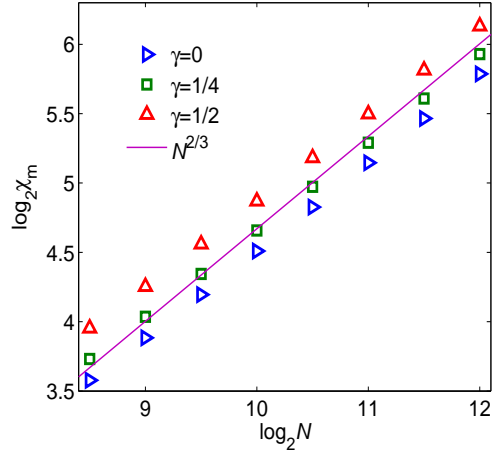}
\end{center}
\caption{Maximum susceptibility $\protect\chi _{m}$ as a function of system
size $N$. We can see that the numerical results approach to the solid line
with slope $2/3$ as the system size increases.}
\label{log_n_log_sus}
\end{figure}
Immediately, one can obtain the asymptotic form of all the spin expectation
values%
\begin{align}
\frac{2\langle S_{z}\rangle }{N}\bigg{|}_{h=1}& \sim 1+\frac{1}{N}+\frac{%
a_{z}^{\left( 0\right) }}{N^{2/3}},  \notag \\
\frac{4\langle S_{x}^{2}\rangle }{N^{2}}\bigg{|}_{h=1}& \sim \frac{%
a_{xx}^{\left( 0\right) }}{N^{2/3}},  \notag \\
\frac{4\langle S_{y}^{2}\rangle }{N^{2}}\bigg{|}_{h=1}& \sim \frac{%
a_{yy}^{\left( 0\right) }}{N^{4/3}},  \notag \\
\frac{4\langle S_{z}^{2}\rangle }{N^{2}}\bigg{|}_{h=1}& \sim 1+\frac{2}{N}+%
\frac{a_{zz}^{\left( 0\right) }}{N^{2/3}}.
\end{align}%
where $a_{\xi }^{\left( 0\right) }$'s ($\xi =z,xx,yy,zz$) are all constants
depending on $\gamma $. Then take the first-order derivatives of Eq.~(\ref%
{spinval}) with $h$, one could find similar scaling functions with Eq.~(\ref%
{scaling form of the singular part}). Here we also take $2\langle
S_{z}\rangle /N$ for example,%
\begin{figure}[tbp]
\begin{center}
\includegraphics[
height=2.55in, width=2.8in ]{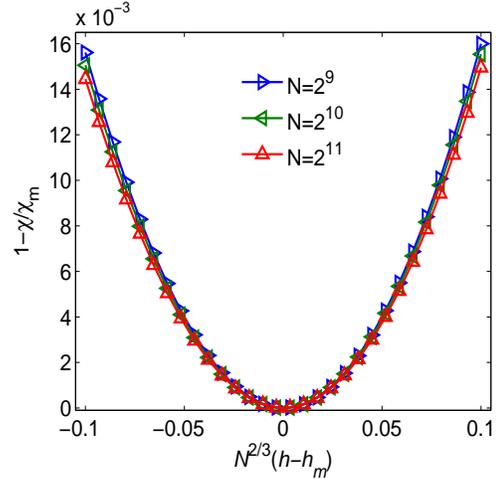}
\end{center}
\caption{Finite size scaling is performed. The susceptibility $\protect\chi $
is considered as a function of the system size $N$ and the parameter $h$,
take the form $N^{\protect\nu }\left( h-h_{m}\right) $. Here the exponent $%
\protect\nu =2/3$ is determined analytically. It's noticed that the data
dose not collapse at one line exactly since the system sizes are not large
enough.}
\label{scabh_sus_n2_3}
\end{figure}
\begin{equation}
\left( \frac{\partial }{\partial h}\frac{2\langle S_{z}\rangle }{N}\right) ^{%
\text{sing}}\simeq \frac{1}{NG\left( h,\gamma \right) ^{3/2}}\mathscr{G}%
_{S_z}\left[ NG\left( h,\gamma \right) ^{3/2},\gamma \right] ,
\label{sing form of nth order diff}
\end{equation}%
where $\mathscr{G}_{\Phi}$ is a scaling function for the derivatives of spin
expectation values, and then we find the finite size scaling%
\begin{equation}
\frac{\partial }{\partial h}\frac{2\langle S_{z}\rangle }{N}\bigg{|}%
_{h=1}\sim a_{z}^{\left( 1\right) }.
\end{equation}%
The scaling form of the other derivatives are%
\begin{align}
\frac{\partial }{\partial h}\frac{4\langle S_{x}^{2}\rangle }{N^{2}}\bigg{|}%
_{h=1}& \sim a_{xx}^{\left( 1\right) },  \notag \\
\frac{\partial }{\partial h}\frac{4\langle S_{y}^{2}\rangle }{N^{2}}\bigg{|}%
_{h=1}& \sim \frac{a_{yy}^{\left( 1\right) }}{N^{2/3}},  \notag \\
\frac{\partial }{\partial h}\frac{4\langle S_{z}^{2}\rangle }{N^{2}}\bigg{|}%
_{h=1}& \sim a_{zz}^{\left( 1\right) },
\end{align}%
where $a_{\xi }^{\left( 1\right) }$'s ($\xi =z,xx,yy,zz$) are
constants depending on $\gamma $. As we can see that, except for
$4\langle S_{y}^{2}\rangle /N^{2}$, the other first-order
derivatives are all independent of $N$. Then with the help of
Eq.~(\ref{gen_sus}), we find that
the maximum RFS $\chi _{m}\equiv \chi \left( h_{m},N,\gamma \right) $ is%
\begin{equation}
\chi _{m}\sim -\frac{\left( a_{zz}^{\left( 1\right) }\right) ^{2}N}{%
a_{zz}^{\left( 0\right) }N^{1/3}+2},
\end{equation}%
for large $N$, and here we just present the divergent term. It's
noticed that $a_{zz}^{\left( 0\right) }$ should be less than $-2$ to
ensure the matrix element $y>0$, thus $\chi _{m}>0$. Then we have
\begin{equation}
\ln \chi _{m}=A_{N}\ln N+\text{const.}~,  \label{sus_n}
\end{equation}%
where the constant only depends on $\gamma $ and the scaling exponent $A_{N}$
approaches to $2/3$ as $N$ increases, which is verified numerically, and $%
A_{N}=2/3$ in thermodynamic limit. The numerical comparisons are shown in
Fig.~(\ref{log_n_log_sus}). While in the broken symmetric phase ($0<h<1$),
we can derive the same scaling exponents \cite{dusuel:224420}. However, for
global fidelity susceptibility, the scaling exponent is $9/7$ \cite%
{Kwok:arxiv/0710.2581v1}.

Then if we cancel $N$ in Eq.~(\ref{sing form of nth order diff}), with
similar steps, we can get the relation between the susceptibility $\chi $
and $\eta =h-h_{c}$ in thermodynamic limit,%
\begin{equation}
\ln \chi \left( h,\gamma \right) =A_{h}\ln \left\vert h-h_{c}\right\vert +%
\text{const.}~,
\end{equation}%
where $A_{h}$ approaches to $-1$ as $h$ goes to $h_{c}$, and the constant
depends on $\gamma $. Therefore we could take the form of the susceptibility
for finite size as
\begin{equation}
\chi \left( h,N,\gamma \right) =\frac{A}{N^{-2/3}+B\left( h-h_{m}\right) }.
\label{sus h}
\end{equation}
To study the critical behavior around the phase transition point, we could
perform the finite scaling analysis. According to the scaling ansatz \cite%
{BarberPhaseTran}, the susceptibility is a function of $N^{\nu }\left(
h-h_{m}\right) $. In the case of logarithmic divergence, it behaves as $\chi
\left( h_{m},N\right) /\chi \left( h,N\right) \sim Q\left[ N^{v}\left(
h-h_{m}\right) \right] $, where the function $Q(x)\approx \ln x$ for large $%
x $ is universal and does not depend on system size $N$. Hence with Eqs.~(%
\ref{sus_n}) and (\ref{sus h}), we determine the exponent $v=2/3$,
which is confirmed numerically, as shown in
Fig.~(\ref{scabh_sus_n2_3}). However, the curves for different
system sizes does not collapse to a single one exactly, since the
system sizes are not large enough.~\newline

\section{Conclusion}

In summary, we have investigated the RFS in the second order quantum
phase transition of the LMG model. For the case that $\rho$ is block-diagonal in $%
2\times2$ matrices, we derive a general formula for RFS. Then with
the CUTs and the scaling ansatz, the critical exponents, including
the finite-size scaling exponents of the RFS are obtained
analytically, and confirmed numerically. Our results show that, the
RFS undergoes singularity around the critical point, indicating that
the RFS can be used to characterize the QPTs. And it's suggested
that we can extract information of the QPTs only from the fidelity
of a subsystem, without probing the global system, which is of
practical significance in experiments. It is also interesting to
study finite-size scaling of RFS in other models such as quantum
Ising model, which is under consideration.

\section{Acknowledgements}
We are indebted to Shi-Jian Gu, C. P. Sun and Z. W. Zhou for
fruitful and valuable discussions. The work was supported by the
Program for New Century Excellent Talents in University (NCET), the
NSFC with grant nos. 90503003, the State Key Program for Basic
Research of China with grant nos. 2006CB921206, the Specialized
Research Fund for the Doctoral Program of Higher Education with
grant No.20050335087.

\end{document}